\documentclass[journal=apchd5,manuscript=article]{achemso}
\usepackage{graphicx} % Required for inserting images
\usepackage{xr}
\author{Lee Grimberg}
\affiliation[Ben-Gurion University]
{Department of Materials Engineering, Ben-Gurion University of the Negev, Be'er Sheva 84105, Israel}
\author{Svyatoslav Kostyukovets}
\affiliation[Ben-Gurion University]
{Department of Materials Engineering, Ben-Gurion University of the Negev, Be'er Sheva 84105, Israel}
\author{Moshe G. Harats}
\affiliation[Ben-Gurion University]
{Department of Materials Engineering, Ben-Gurion University of the Negev, Be'er Sheva 84105, Israel}
\email{mharats@bgu.ac.il}
\title{Angular Emission Properties of Strained Transition-Metal Dichalcogenides}

\begin{document}
\begin{abstract}
Monolayers of transition-metal dichalcogenides have shown that uniaxial strain changes both the photoluminescence emission energy and intensity. The changes are attributed to the band-structure evolution under tensile strain where both the bandgap decreases and a direct-to-indirect transition occurs. This was shown for relatively high strains, whereas this is not the case at low strain values $<1\%$ in which in this work, we observe the erratic dependency of the photoluminescence intensity at low strain values as a function of strain. We find that the dominant physical property is the dependence of the optical-dipole emission on the curvature of the substrate. We validate the behavior of the photoluminescence intensity with experimental angular emission spectroscopy (k-space imaging). These findings are supported by Finite-Difference Time-Domain simulations, in agreement with the experimental data. Our findings present the importance of choosing the right substrate for flexible devices based on transition-metal dichalcogenides.
\end{abstract}
% \maketitle

\section{Introduction}
Transition-metal dichalcogenides (TMDs) are considered as prime candidates for flexible devices such as solar-cells \cite{Feng2012}, transistors \cite{Lee2014,Kang2025StrainElectrodes}, and quantum light emitters \cite{Iff2019Strain-TunableMonolayers,Paralikis2024TailoringEngineering,Grosso2017}. This is due to their elastic properties - typical TMDs have a Young's modulus of $\approx250GPa$ \cite{Liu2016,Zhang2016,Castellanos-Gomez2012} and can sustain strain above $10\%$ as well. In many works that present the flexibility of TMDs devices, authors show that the devices such as gas sensors \cite{Goswami2022RecentSensing,Kumar2020TransitionSensors} and photodiodes \cite{Vasconcelos2025Strain-EngineeredSpectrometry} can withstand thousands of bending cycles without any deterioration in the performance of the devices. Yet, when one desires to use a flexible device, it is advantageous to fully understand the operation \emph{during} the bending cycle itself.

In the pioneering works that have shown the role of strain in monolayers TMDs, the emphasis was on the redshift of the exciton emission line as a function of tensile strain. This was shown both for uniaxial \cite{Conley2013,Niehues2018StrainSemiconductors}, biaxial \cite{Lloyd2016}, and non-uniform strain \cite{Harats2020DynamicsWS2}, in the high strain ($>1\%$) regime. Moreover, the decrease in the intensity of the emission was a main feature in these works and was shown to be a result of the closing of the bandgap - the transition from a direct bandgap to an indirect one \cite{Conley2013,Lloyd2016}. Nevertheless, the role of the substrate on which the TMDs monolayer was supported was not taken into account.

In this work, we incorporate spectrally-resolved k-space (or back-focal-plane) imaging to elucidate the angular emission properties of monolayer TMDs as a function of applied uniaxial tensile strain at the low strain limit ($<1\%$) - the usual operational strain of commercial devices. We implement k-space imaging to reveal 2 complementary findings: (\emph{i}) the optical dipole orientation in a monolayer TMD is in-plane (IP) and does not change under small uniaxial strain and (\emph{ii}) the intensity of the emission is highly dependent on strain without any general trend related to the curvature of the supporting substrate. We assess our findings with both numerical simulations and a theoretical model that show a good agreement with the experimental data. 

\section{Experimental setup and results}
\begin{figure}
    \centering
    \includegraphics[width=0.9\linewidth]{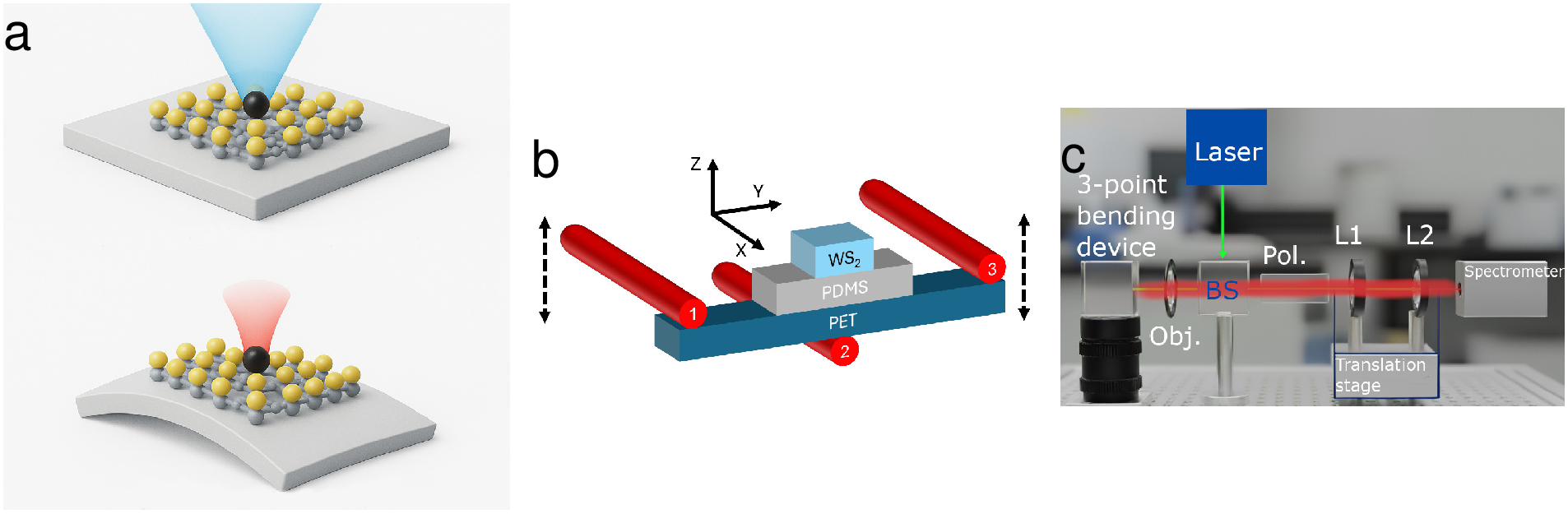}
    \caption{(a) Description of the experiment. Top - an unstrained sample emits light into the numerical aperture of a lens. Bottom -  when strained, the emission redshifts and the light cone slightly changes, influencing the brightness of the emission. (b) A schematic illustration of the 3-point bending device. Note that the PET is strained directly and not the PDMS. The outer posts numbered 1,3 move up and down to strain the device while the middle post number 2 (where the sample is centered) does not move as it stays at the focus of the illuminating objective. (c) The experimental setup. BS - beam splitter. L1 - the k-space (Bertrand) lens that flips for PL acquisition. L2 - the imaging lens. }
    \label{fig:fig1}
\end{figure}

In this work, we used a monolayer of a typical TMD, $WS_2$. The monolayer was mechanically exfoliated on a commercial PDMS stamp. The PDMS stamp with the $WS_2$ monolayer (Fig. S1) was transferred onto a PET substrate creating a $WS_2/PDMS/PET$ device (see Fig. \ref{fig:fig1}b).

Unlike other experiments where the monolayer is usually transferred directly to the PET substrate without the PDMS stamp, we found in our initial investigations that the PET emits light in the same optical regime as the photoluminescence (PL) of $WS_2$. In the case of real space PL measurements, background subtraction suffices, while it is not the case in k-space measurements where the light emission from the PET is comparable to the emission from the $WS_2$ monolayer. As the emission from PDMS was found to be negligible compared to PET, we chose the layered sample of $WS_2/PDMS/PET$ for this work. 

To apply uniaxial strain in a controlled manner, we implemented a 3-point bending device following Ref. [\citenum{Cakroglu2023AnMaterials}]. The layered $WS_2/PDMS/PET$ sample is mounted on the bending device where we carefully align the monolayer at the center of the central post among the 3 posts of the bending device (see Fig. \ref{fig:fig1}b).

The optical setup is depicted in Fig. \ref{fig:fig1}c. The illumination and PL of the sample is done in a backscattering configuration where the PL is analyzed by a high-resolution spectrometer. The k-space is imaged with a Bertrand-lens that images the back-focal plane of the illuminating objective onto the entrance slit of the spectrometer \cite{Harats2014,Livneh2016,Livneh2015EfficientNanoantenna}. The sample can be illuminated either with a 532nm CW diode laser or by a white light source used for imaging the sample. The sample is illuminated with a high numerical aperture (NA=0.75) objective. The high NA is necessary for covering a large portion of the angular emission from the sample.

\begin{figure}
    \centering
    \includegraphics[width=\linewidth]{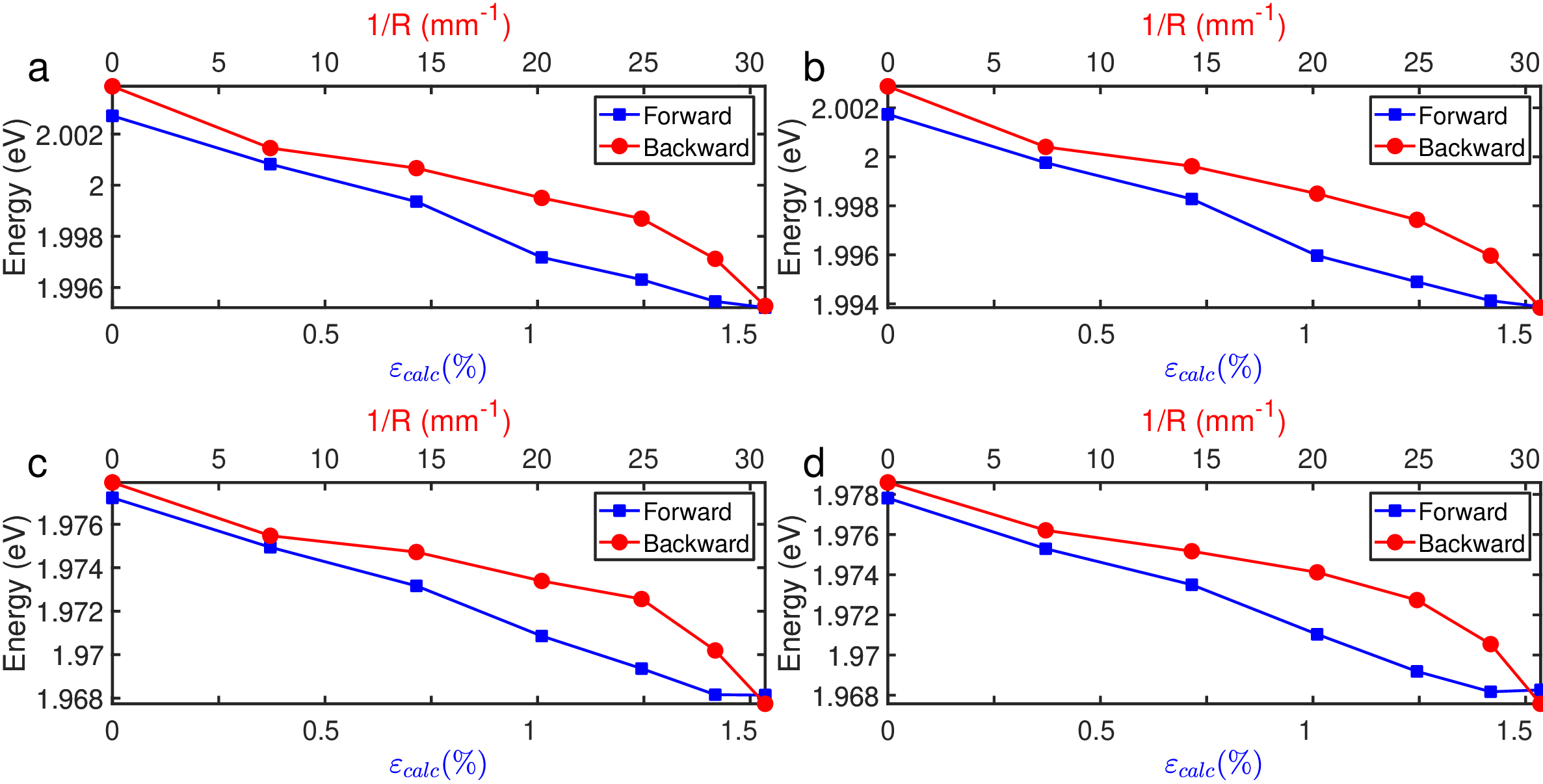}
    \caption{The hysteresis of the energies of the excitons and the trions for both polarizations. The blue curves correspond to the forward direction (increasing strain) and the red to the backward direction (decreasing strain). The bottom x-axis (blue) corresponds to the classical calculated strain of the 3-point bending device and the top x-axis (red) corresponds to the inverse curvature. (a) TE exciton energy. (b) TM exciton energy. (c) TE trion energy. (d) TM trion energy.}
    \label{fig:fig3_hysteresis}
\end{figure}

The experimental setup allows measurements both the samples PL at different polarizations (TE and TM) and the corresponding k-space at the same polarization. The strain is controlled by the movement of the outer posts of the bending device. %Although the strain can be calculated directly from the geometrical parameters of the bending device \cite{Conley2013,Niehues2018StrainSemiconductors}, we observe that the shift of the exciton energy does not correspond to this strain value (see Fig. S2). This is due to PDMS which, being a viscoelastic material, does not fully transfer the strain from the PET to the $WS_2$ monolayer and shows a hysteretic behavior (see Fig. \ref{fig:S6_hysteresis}). This is beneficial for the purpose of this work as we are able to remain in the limit of low strain values $\varepsilon<0.2\%$.
We begin with measuring the PL from the monolayer as a function of the curvature of the PET substrate (see Fig. S2). The elastic response of the monolayer is related to the strain transfer from the strained PET, through the PDMS (that is not strained directly - see Fig. \ref{fig:fig1}b) which is a viscoelastic material, onto the $WS_2$ monolayer. The viscoelastic behavior is shown in Fig. \ref{fig:fig3_hysteresis} where the exciton and trion energies show a hysteretic behavior as a function of increasing and decreasing strain. The lower x-axis corresponds to the calculated strain of the PET in a 3-point bending device:

\begin{equation}
    \varepsilon_{calc}=\frac{6Dt}{L^2}
\end{equation}\label{eq:eps_calc}
where $D$ is the deflection of the 3-point bending device, $t=125\mu m$ is the thickness of the PET, and $L=24mm$ is the distance between the outer posts. It is evident that on top of the hysteresis, the strain is not fully transferred as we do not observe the redshift-strain ratio of $50 meV/\%$ reported in various works \cite{Niehues2018StrainSemiconductors,Conley2013,Lloyd2016,Harats2020DynamicsWS2}. Therefore, we benefit from the viscoelasticity of the PDMS as we remain in the low strain regime of $\varepsilon<0.2\%$ and the real (not calculated) strain values presented in the rest of the paper are extracted from the fitted exciton energies from the measured spectra. The top x-axis corresponds to the inverse of the curvature (as the curvature is infinite for the case of no strain it is more convenient to introduce the inverse curvature):

\begin{equation}\label{eq:inv_curvature}
    \frac{1}{R}=\frac{4d}{ABC}
\end{equation}
where $A=C$ are the distance between the outer points of the bending device and the central point, $B$ is the distance between the outer points, and $d$ is the area of the triangle that they define.

\begin{figure}
    \centering
    \includegraphics[width=\linewidth]{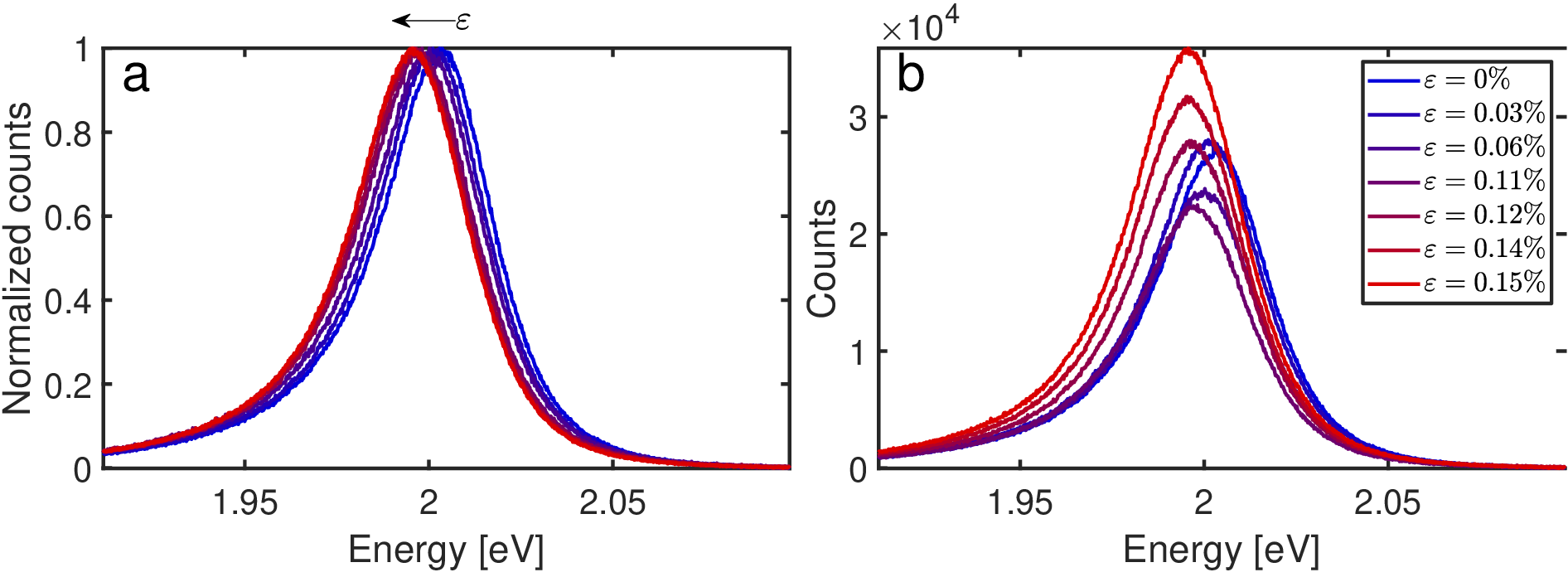}
    \caption{(a) The normalized PL as a function of strain. The excitonic peak red-shifts as a function of strain as previously reported \cite{Lloyd2016,Conley2013,Harats2020DynamicsWS2}. (b) The non-normalized PL from (a) showing fluctuations in the PL intensity without a clear trend. }
    \label{fig:fig3_PL}
\end{figure}

Before presenting the dependency of the PL intensity on the strain, we note that the starting point, $\varepsilon=0\%$, is not exactly stress-free as the PET substrate is strained by a very small value due to the experimental conditions. To ensure that the PET does not slide, we exert minimal force from the outer posts of the bending device which implies that some initial (minimal) strain is applied also at our starting point. Therefore, all the strain values presented are relative to $\varepsilon=0\%$, meaning that between different samples and experiment cycles, the starting point may be different. In addition, the polarization shown in this manuscript is in TE polarization unless stated otherwise (the corresponding TM polarization, that  shows $\sim 35\%$ less intensity, is presented in Fig. S3). In Fig. \ref{fig:fig3_PL}a the normalized PL shows a redshift in the peak energy of the PL (corresponding to the A-exciton peak), in agreement with numerous reports on the energy shift of the PL as a function of strain \cite{Conley2013,Lloyd2016,Harats2020DynamicsWS2,Kovalchuk2020NeutralWS2}. Surprisingly, when we look at the non-normalized PL (shown in Fig. \ref{fig:fig3_PL}b), there is no visible trend of the intensity of the PL, whereas it seems to fluctuate. This is in contradiction with the direct-to-indirect transition as a function of strain that has been shown to decrease the PL monotonically \cite{Conley2013,Lloyd2016}. 

\begin{figure}
    \centering
    \includegraphics[width=1\linewidth]{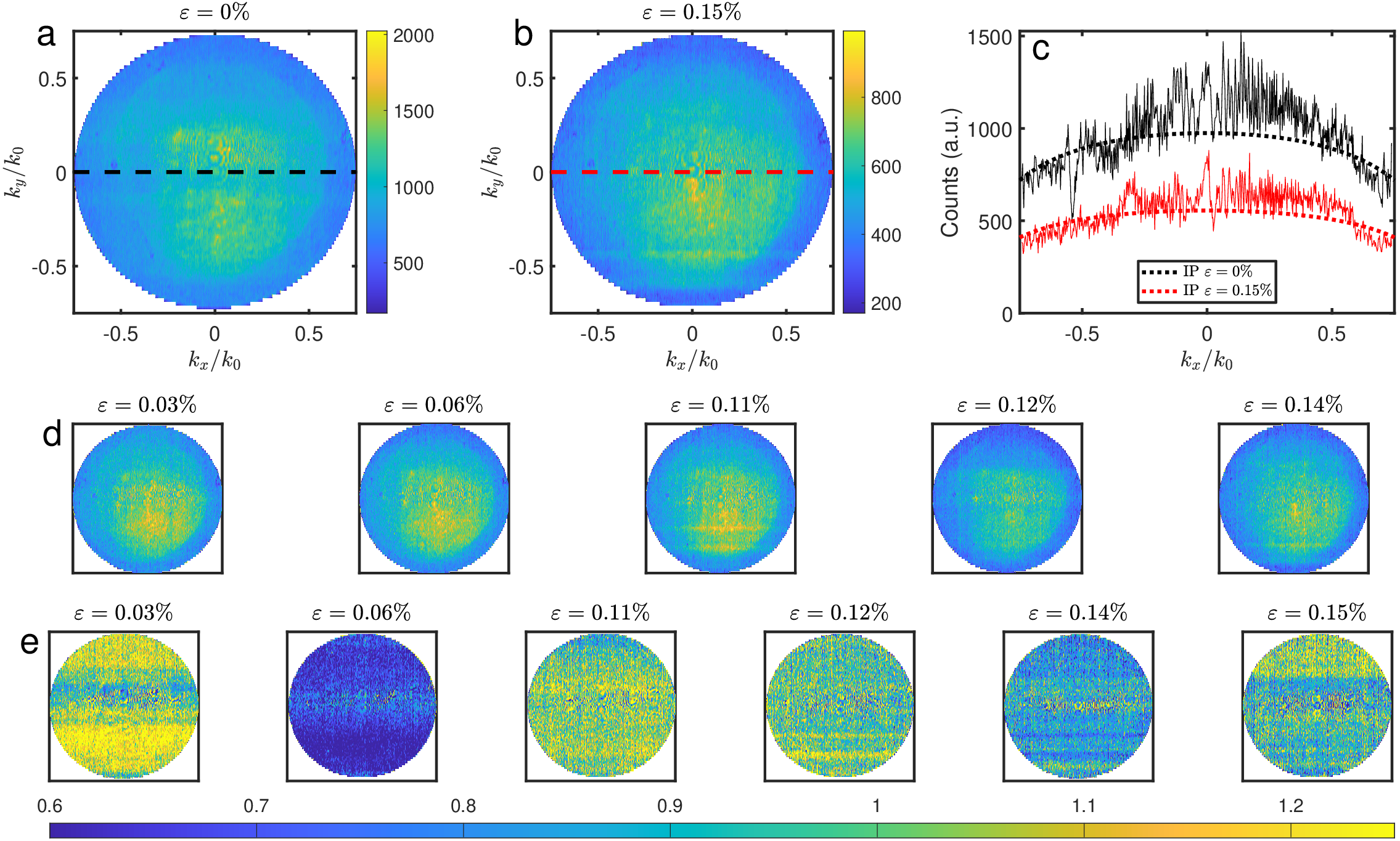}
    \caption{Angular (k-space) emission spectrally resolved at $618nm$ for TE polarization. (a) The k-space at the lowest strain $\varepsilon=0\%$. (b) The k-space at the highest strain $\varepsilon=0.15\%$. (c) The angular emission at $k_y=0$ corresponding to the cases of no strain (dashed black line in (a)) and maximum strain (dashed red line in (b)). The calculations of IP dipole emission are presented as dotted curves and show that the emission consists of IP dipoles without any change due to strain. (d) Same as (a) and (b) for intermediate strain values. The colorbars and axis are omitted for clarity. (e) K-space emission normalized to the previous strain value. The fringes are changing along the $k_y$ direction. The colorbar at the bottom is common for all the normalized k-space.}
    \label{fig:fig2}
\end{figure}

To elucidate the peculiar behavior of the PL intensity, we implemented spectrally and polarized resolved k-space imaging (SPRKI). With the SPRKI method we were able to image the back-focal plane of the illuminating objective onto a spectrometer and therefore, obtain the spectrally-resolved k-space (or angular emission) PL from the sample \cite{Harats2014,Livneh2016,Livneh2015EfficientNanoantenna}. In Fig. \ref{fig:fig2} we present the angular emission at $618nm$ (corresponding to the energy of the exciton at $\varepsilon=0\%$) both for the sample when no strain (Fig. \ref{fig:fig2}a) and the maximal strain (Fig. \ref{fig:fig2}b) is applied. First, the angular emission is characteristic of an IP optical dipole as reported previously for the case of no strain \cite{Schuller2013OrientationNanomaterials,Brotons-Gisbert2019Out-of-planeSelenide} (see Fig. \ref{fig:fig2}c). The dashed curves in Fig. \ref{fig:fig2}c (superimposed with the experimental data) are calculated as an ensemble of IP dipoles in proximity to a flat PDMS substrate \cite{Novotny2012PrinciplesNano-Optics,Lukosz1979LightOrientation}. As the experimental data follows the IP dipole angular emission pattern, we confirm that the optical dipole orientation does not change under applied uniaxial strain. 

In Fig. \ref{fig:fig2}d we present the angular emission for all strain values, where fringes are observed along the $k_y$ which corresponds to the axis of the bending of the sample (see Fig. \ref{fig:fig1}b). These fringes appear also for all strain values, and are highlighted when we normalize the angular emission of each strain value to its precedent (see Fig. \ref{fig:fig2}e). The appearance of the fringes at the $\varepsilon=0\%$ is due to the experimental conditions that apply a small amount of strain, as stated above. Looking deeper into the data presented in Fig. \ref{fig:fig2}e, we observe not only the shift in the location of the fringes in the $k_y$ direction between subsequent strain values, but also in the amplitude. This is another manifestation of the erratic behavior of the PL as a function of strain that is visible in the k-space measurements as well.

\section{Discussion}
As these fringes appear also in TM polarization and at the trion energy at $627nm$ (see Figs S4,S5,S6), we investigated our system using Finite-Difference Time-Domain (FDTD) simulations (MEEP) \cite{Oskooi2010Meep:Method}. In our simulation cell we located the optical dipole at a location of $3.5nm$ from the PDMS interface (see Fig. S7). The PDMS was assumed flat for the case of $\varepsilon=0\%$, and curved for the other strain values for which the curvature was calculated using Eq. \ref{eq:inv_curvature}.
% \begin{equation}\label{eq:curvature}
%     R=\frac{ABC}{4D}
% \end{equation}
% where $A=C$ are the distance between the outer points of the bending device and the central point, $B$ is the distance between the outer points, and $D$ is the area of the triangle that they define.

\begin{figure}
    \centering
    \includegraphics[width=1\linewidth]{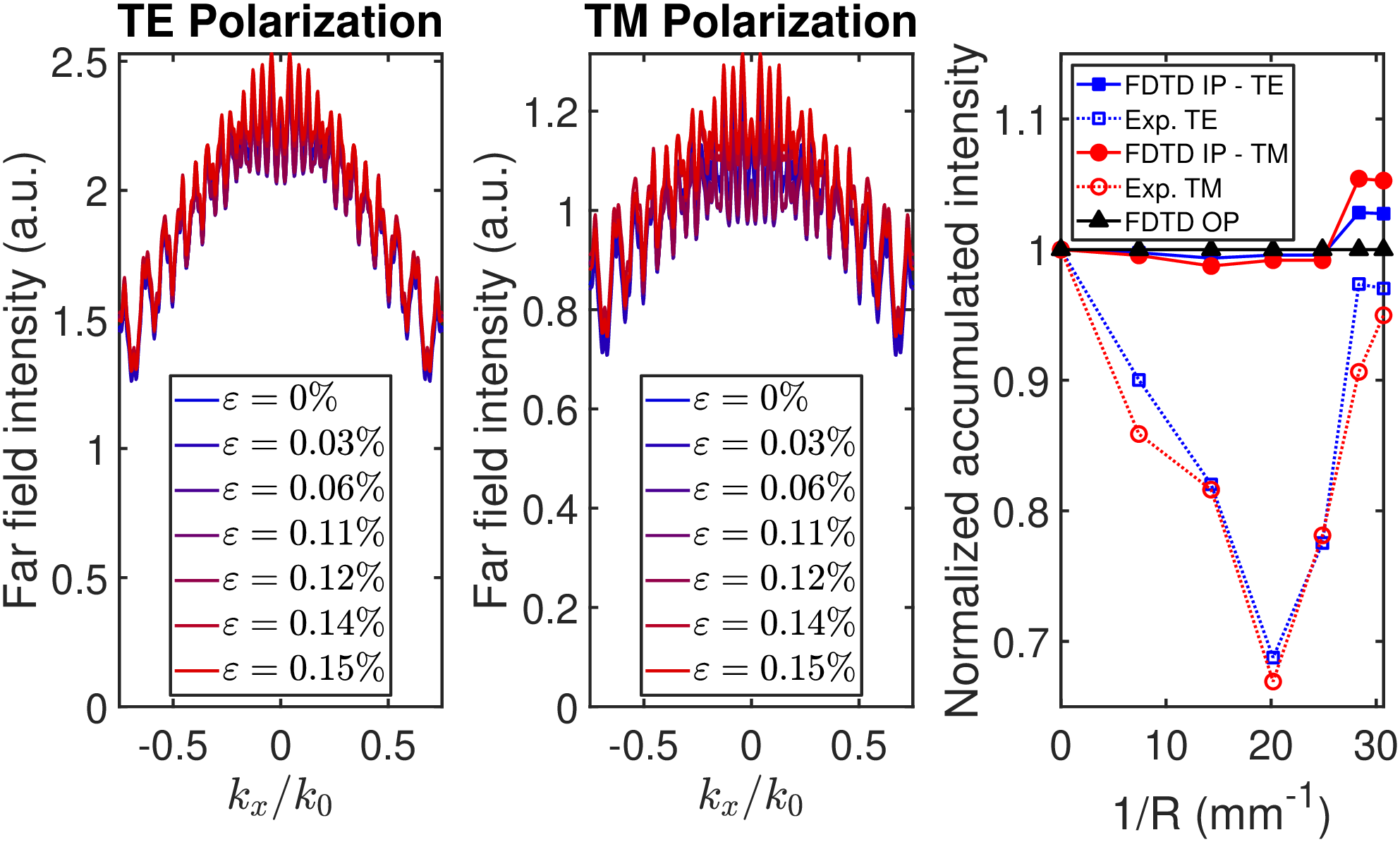}
    \caption{(a) FDTD far field pattern for an IP dipole in TE polarization. The variations between different strain values is visible. (b) Same as (a) for TM polarization. Note the lower intensity compared to TE polarization. (c) The normalized far-field intensities (FDTD and experimental). The similarity in the trend between the IP simulation results and the experimental values is striking. The OP dipole is negligible and indeed does not change under strain.}
    \label{fig:FDTD}
\end{figure}

In Fig. \ref{fig:FDTD} we show the simulated far-field emission from an IP dipole located $3.5nm$ from the PDMS substrate at 2 different configurations that correspond to the TE (Fig. \ref{fig:FDTD}a) and TM (Fig. \ref{fig:FDTD}b) polarizations in the experiments. The out-of-plane (OP) dipole does not change at all as a function of the strain (see Fig. S7). A summary of the OP dipole is shown in Fig. \ref{fig:FDTD}c (solid lines) where we compare the normalized integrated far-field (essentially the integration of the curves in Fig. \ref{fig:FDTD}a-b normalized to the integrated far-field for $\varepsilon=0\%$) for the different dipole orientations as a function of strain. The dashed lines present the normalized experimental intensity at $618nm$ as a function of strain (for both polarizations) and we observe that the trend follows remarkably the FDTD simulation results. The fact that at the largest strains the normalized experimental intensities do not exceed 1 unlike the FDTD simulations results, is the manifestation of the direct-to-indirect transition that is convoluted with our findings. Indeed, the FDTD simulations confirm our experimental observations. 

What is the physical mechanism behind the different PL intensities as a function of curvature / strain? To understand our experimental and simulation results, we return to the textbook solution of dipole emission close to a dielectric interface \cite{Lukosz1979LightOrientation,Novotny2012PrinciplesNano-Optics}. In the static limit, the solution can be found by the image-dipole approximation. In this approximation, for the IP dipoles (see section 10.10 in Ref. [\citenum{Novotny2012PrinciplesNano-Optics}]), the electric fields that the dipole produces in the backscattered region are a superposition of the real and image dipole electric fields, where the image dipole fields are multiplied by the Fresnel coefficient $r_{s,p}$.
% \begin{equation}\label{eq:fresnel}
%     r^p=\frac{n_2-n_1}{n_2+n_1}
% \end{equation}
% where $n_{1,2}$ are the refractive indices of air and PDMS. 
In our case, the Fresnel reflection coefficients depend on the curvature $R$ (Eq. \ref{eq:inv_curvature}) \cite{Hentschel2002FresnelMicroresonators,Luhn2019AnalyticalInterfaces} or, with the correct curvature-strain relation, to the strain $\varepsilon$. This implies that changes in the reflection coefficients as a function of strain correspond to changes in the total PL intensities as observed in this work. Moreover, the interference between the real and image dipole changes as a function of the curvature which explains the changes in the fringes location. Although the changes in the reflection coefficients $r_{s,p}(R)$ are relatively negligible in our case (see Refs. [\citenum{Luhn2019AnalyticalInterfaces,Hentschel2002FresnelMicroresonators}] that discuss the case for microresonators), the fact that we are able to observe the changes in our SPRKI setup shows how this experimental technique is highly sensitive to the effects of uniaxial strain (or curvature) on the optical properties of strained TMDs.

\section{Conclusions}
In this work, we implemented k-space spectroscopy to resolve both the angular and the optical dipole nature of a monolayer of a prototype TMD - $WS_2$. By using the SPRKI technique we were able to confirm that although the optical dipole of a monolayer TMD does not change under strain and stays mainly IP, the PL intensities change in an unpredictable manner at low strain values. This contradiction to the known direct-to-indirect transition which implies that the PL intensity should decrease monotonically as a function of strain, is revealed by our SPRKI measurements. The observed fringes, along the direction of the bending of the monolayer, is a result of the different Fresnel reflection coefficients that depend on the curvature, both in TE and TM polarizations. Our results emphasize the challenges in achieving quantitative and precise PL yield from light-emitting devices based on strained TMDs monolayers with a transparent flexible substrate, as the fluctuations in the device PL intensity is altered due to the curvature of the substrate. 

\section{Supporting Information}
Supporting information: Sample fabrication, additional experimental results, FDTD calculations details.

\begin{acknowledgement}
We thank funding from the Israel Science Foundation (ISF) 492/25.
The authors thank Yonatan Sivan for fruitful discussions and Andres Castellanos-Gomez for providing the files for the bending device \cite{Cakroglu2023AnMaterials}.

\end{acknowledgement}

\bibliography{references}

\providecommand{\latin}[1]{#1}
\makeatletter
\providecommand{\doi}
  {\begingroup\let\do\@makeother\dospecials
  \catcode`\{=1 \catcode`\}=2 \doi@aux}
\providecommand{\doi@aux}[1]{\endgroup\texttt{#1}}
\makeatother
\providecommand*\mcitethebibliography{\thebibliography}
\csname @ifundefined\endcsname{endmcitethebibliography}  {\let\endmcitethebibliography\endthebibliography}{}
\begin{mcitethebibliography}{29}
\providecommand*\natexlab[1]{#1}
\providecommand*\mciteSetBstSublistMode[1]{}
\providecommand*\mciteSetBstMaxWidthForm[2]{}
\providecommand*\mciteBstWouldAddEndPuncttrue
  {\def\EndOfBibitem{\unskip.}}
\providecommand*\mciteBstWouldAddEndPunctfalse
  {\let\EndOfBibitem\relax}
\providecommand*\mciteSetBstMidEndSepPunct[3]{}
\providecommand*\mciteSetBstSublistLabelBeginEnd[3]{}
\providecommand*\EndOfBibitem{}
\mciteSetBstSublistMode{f}
\mciteSetBstMaxWidthForm{subitem}{(\alph{mcitesubitemcount})}
\mciteSetBstSublistLabelBeginEnd
  {\mcitemaxwidthsubitemform\space}
  {\relax}
  {\relax}

\bibitem[Feng \latin{et~al.}(2012)Feng, Qian, Huang, and Li]{Feng2012}
Feng,~J.; Qian,~X.; Huang,~C.-W.; Li,~J. {Strain-engineered artificial atom as a broad-spectrum solar energy funnel}. \emph{Nature Photonics} \textbf{2012}, \emph{6}, 866--872\relax
\mciteBstWouldAddEndPuncttrue
\mciteSetBstMidEndSepPunct{\mcitedefaultmidpunct}
{\mcitedefaultendpunct}{\mcitedefaultseppunct}\relax
\EndOfBibitem
\bibitem[Lee \latin{et~al.}(2014)Lee, Lee, van~der Zande, Chen, Li, Han, Cui, Arefe, Nuckolls, Heinz, Guo, Hone, and Kim]{Lee2014}
Lee,~C.-H.; Lee,~G.-H.; van~der Zande,~A.~M.; Chen,~W.; Li,~Y.; Han,~M.; Cui,~X.; Arefe,~G.; Nuckolls,~C.; Heinz,~T.~F.; Guo,~J.; Hone,~J.; Kim,~P. {Atomically thin p–n junctions with van der Waals heterointerfaces}. \emph{Nature Nanotechnology} \textbf{2014}, \emph{9}, 676--681\relax
\mciteBstWouldAddEndPuncttrue
\mciteSetBstMidEndSepPunct{\mcitedefaultmidpunct}
{\mcitedefaultendpunct}{\mcitedefaultseppunct}\relax
\EndOfBibitem
\bibitem[Kang \latin{et~al.}(2025)Kang, Ji, Park, Kim, Lee, Lim, Park, Kim, Park, Hong, Yoo, and Choi]{Kang2025StrainElectrodes}
Kang,~M.; Ji,~J.; Park,~S.; Kim,~S.-B.; Lee,~I.; Lim,~H.; Park,~C.; Kim,~S.; Park,~H.; Hong,~W.; Yoo,~S.; Choi,~S.-Y. {Strain Engineering of a Dual-Gate Structure for Highly Flexible and Transparent MoS2 Thin-Film Transistors with Graphene Electrodes}. \emph{ACS Nano} \textbf{2025}, \relax
\mciteBstWouldAddEndPunctfalse
\mciteSetBstMidEndSepPunct{\mcitedefaultmidpunct}
{}{\mcitedefaultseppunct}\relax
\EndOfBibitem
\bibitem[Iff \latin{et~al.}(2019)Iff, Tedeschi, Mart{\'{i}}n-S{\'{a}}nchez, Mocza{\l}a-Dusanowska, Tongay, Yumigeta, Taboada-Guti{\'{e}}rrez, Savaresi, Rastelli, Alonso-Gonz{\'{a}}lez, H{\"{o}}fling, Trotta, and Schneider]{Iff2019Strain-TunableMonolayers}
Iff,~O.; Tedeschi,~D.; Mart{\'{i}}n-S{\'{a}}nchez,~J.; Mocza{\l}a-Dusanowska,~M.; Tongay,~S.; Yumigeta,~K.; Taboada-Guti{\'{e}}rrez,~J.; Savaresi,~M.; Rastelli,~A.; Alonso-Gonz{\'{a}}lez,~P.; H{\"{o}}fling,~S.; Trotta,~R.; Schneider,~C. {Strain-Tunable Single Photon Sources in WSe2 Monolayers}. \emph{Nano Letters} \textbf{2019}, \emph{19}, 6931--6936\relax
\mciteBstWouldAddEndPuncttrue
\mciteSetBstMidEndSepPunct{\mcitedefaultmidpunct}
{\mcitedefaultendpunct}{\mcitedefaultseppunct}\relax
\EndOfBibitem
\bibitem[Paralikis \latin{et~al.}(2024)Paralikis, Piccinini, Madigawa, Metuh, Vannucci, Gregersen, and Munkhbat]{Paralikis2024TailoringEngineering}
Paralikis,~A.; Piccinini,~C.; Madigawa,~A.~A.; Metuh,~P.; Vannucci,~L.; Gregersen,~N.; Munkhbat,~B. {Tailoring polarization in WSe2 quantum emitters through deterministic strain engineering}. \emph{npj 2D Materials and Applications 2024 8:1} \textbf{2024}, \emph{8}, 1--8\relax
\mciteBstWouldAddEndPuncttrue
\mciteSetBstMidEndSepPunct{\mcitedefaultmidpunct}
{\mcitedefaultendpunct}{\mcitedefaultseppunct}\relax
\EndOfBibitem
\bibitem[Grosso \latin{et~al.}(2017)Grosso, Moon, Lienhard, Ali, Efetov, Furchi, Jarillo-Herrero, Ford, Aharonovich, and Englund]{Grosso2017}
Grosso,~G.; Moon,~H.; Lienhard,~B.; Ali,~S.; Efetov,~D.~K.; Furchi,~M.~M.; Jarillo-Herrero,~P.; Ford,~M.~J.; Aharonovich,~I.; Englund,~D. {Tunable and high-purity room temperature single-photon emission from atomic defects in hexagonal boron nitride}. \emph{Nature Communications} \textbf{2017}, \emph{8}\relax
\mciteBstWouldAddEndPuncttrue
\mciteSetBstMidEndSepPunct{\mcitedefaultmidpunct}
{\mcitedefaultendpunct}{\mcitedefaultseppunct}\relax
\EndOfBibitem
\bibitem[Liu and Wu(2016)Liu, and Wu]{Liu2016}
Liu,~K.; Wu,~J. {Mechanical properties of two-dimensional materials and heterostructures}. \emph{Journal of Materials Research} \textbf{2016}, \emph{31}, 832--844\relax
\mciteBstWouldAddEndPuncttrue
\mciteSetBstMidEndSepPunct{\mcitedefaultmidpunct}
{\mcitedefaultendpunct}{\mcitedefaultseppunct}\relax
\EndOfBibitem
\bibitem[Zhang \latin{et~al.}(2016)Zhang, Koutsos, and Cheung]{Zhang2016}
Zhang,~R.; Koutsos,~V.; Cheung,~R. {Elastic properties of suspended multilayer WSe2}. \emph{Applied Physics Letters} \textbf{2016}, \emph{108}, 042104\relax
\mciteBstWouldAddEndPuncttrue
\mciteSetBstMidEndSepPunct{\mcitedefaultmidpunct}
{\mcitedefaultendpunct}{\mcitedefaultseppunct}\relax
\EndOfBibitem
\bibitem[Castellanos-Gomez \latin{et~al.}(2012)Castellanos-Gomez, Poot, Steele, van~der Zant, Agra{\"{i}}t, and Rubio-Bollinger]{Castellanos-Gomez2012}
Castellanos-Gomez,~A.; Poot,~M.; Steele,~G.~A.; van~der Zant,~H. S.~J.; Agra{\"{i}}t,~N.; Rubio-Bollinger,~G. {Elastic Properties of Freely Suspended MoS2 Nanosheets}. \emph{Advanced Materials} \textbf{2012}, \emph{24}, 772--775\relax
\mciteBstWouldAddEndPuncttrue
\mciteSetBstMidEndSepPunct{\mcitedefaultmidpunct}
{\mcitedefaultendpunct}{\mcitedefaultseppunct}\relax
\EndOfBibitem
\bibitem[Goswami and Gupta(2022)Goswami, and Gupta]{Goswami2022RecentSensing}
Goswami,~P.; Gupta,~G. {Recent progress of flexible NO2 and NH3 gas sensors based on transition metal dichalcogenides for room temperature sensing}. \emph{Materials Today Chemistry} \textbf{2022}, \emph{23}, 100726\relax
\mciteBstWouldAddEndPuncttrue
\mciteSetBstMidEndSepPunct{\mcitedefaultmidpunct}
{\mcitedefaultendpunct}{\mcitedefaultseppunct}\relax
\EndOfBibitem
\bibitem[Kumar \latin{et~al.}(2020)Kumar, Goel, Hojamberdiev, and Kumar]{Kumar2020TransitionSensors}
Kumar,~R.; Goel,~N.; Hojamberdiev,~M.; Kumar,~M. {Transition metal dichalcogenides-based flexible gas sensors}. 2020\relax
\mciteBstWouldAddEndPuncttrue
\mciteSetBstMidEndSepPunct{\mcitedefaultmidpunct}
{\mcitedefaultendpunct}{\mcitedefaultseppunct}\relax
\EndOfBibitem
\bibitem[Vasconcelos \latin{et~al.}(2025)Vasconcelos, Vladimirov, Pucher, Puebla, Munuera, Hern{\'{a}}ndez, and Castellanos-Gomez]{Vasconcelos2025Strain-EngineeredSpectrometry}
Vasconcelos,~T.~L.; Vladimirov,~S.~N.; Pucher,~T.; Puebla,~S.; Munuera,~C.; Hern{\'{a}}ndez,~E.~R.; Castellanos-Gomez,~A. {Strain-Engineered Adaptive 2D Photodetectors: A New Approach to Miniaturized Reconstructive Spectrometry}. \emph{Nano Letters} \textbf{2025}, \emph{25}, 11333--11339\relax
\mciteBstWouldAddEndPuncttrue
\mciteSetBstMidEndSepPunct{\mcitedefaultmidpunct}
{\mcitedefaultendpunct}{\mcitedefaultseppunct}\relax
\EndOfBibitem
\bibitem[Conley \latin{et~al.}(2013)Conley, Wang, Ziegler, Haglund, Pantelides, and Bolotin]{Conley2013}
Conley,~H.~J.; Wang,~B.; Ziegler,~J.~I.; Haglund,~R.~F.; Pantelides,~S.~T.; Bolotin,~K.~I. {Bandgap Engineering of Strained Monolayer and Bilayer MoS 2}. \emph{Nano Letters} \textbf{2013}, \emph{13}, 3626--3630\relax
\mciteBstWouldAddEndPuncttrue
\mciteSetBstMidEndSepPunct{\mcitedefaultmidpunct}
{\mcitedefaultendpunct}{\mcitedefaultseppunct}\relax
\EndOfBibitem
\bibitem[Niehues \latin{et~al.}(2018)Niehues, Schmidt, Dr{\"{u}}ppel, Marauhn, Christiansen, Selig, Bergh{\"{a}}user, Wigger, Schneider, Braasch, Koch, Castellanos-Gomez, Kuhn, Knorr, Malic, Rohlfing, Michaelis~de Vasconcellos, and Bratschitsch]{Niehues2018StrainSemiconductors}
Niehues,~I. \latin{et~al.}  {Strain Control of Exciton–Phonon Coupling in Atomically Thin Semiconductors}. \emph{Nano Letters} \textbf{2018}, \emph{18}, 1751--1757\relax
\mciteBstWouldAddEndPuncttrue
\mciteSetBstMidEndSepPunct{\mcitedefaultmidpunct}
{\mcitedefaultendpunct}{\mcitedefaultseppunct}\relax
\EndOfBibitem
\bibitem[Lloyd \latin{et~al.}(2016)Lloyd, Liu, Christopher, Cantley, Wadehra, Kim, Goldberg, Swan, and Bunch]{Lloyd2016}
Lloyd,~D.; Liu,~X.; Christopher,~J.~W.; Cantley,~L.; Wadehra,~A.; Kim,~B.~L.; Goldberg,~B.~B.; Swan,~A.~K.; Bunch,~J.~S. {Band Gap Engineering with Ultralarge Biaxial Strains in Suspended Monolayer MoS 2}. \emph{Nano Letters} \textbf{2016}, \emph{16}, 5836--5841\relax
\mciteBstWouldAddEndPuncttrue
\mciteSetBstMidEndSepPunct{\mcitedefaultmidpunct}
{\mcitedefaultendpunct}{\mcitedefaultseppunct}\relax
\EndOfBibitem
\bibitem[Harats \latin{et~al.}(2020)Harats, Kirchhof, Qiao, Greben, and Bolotin]{Harats2020DynamicsWS2}
Harats,~M.~G.; Kirchhof,~J.~N.; Qiao,~M.; Greben,~K.; Bolotin,~K.~I. {Dynamics and efficient conversion of excitons to trions in non-uniformly strained monolayer WS2}. \emph{Nature Photonics 2020 14:5} \textbf{2020}, \emph{14}, 324--329\relax
\mciteBstWouldAddEndPuncttrue
\mciteSetBstMidEndSepPunct{\mcitedefaultmidpunct}
{\mcitedefaultendpunct}{\mcitedefaultseppunct}\relax
\EndOfBibitem
\bibitem[{\c{C}}akıro{\u{g}}lu \latin{et~al.}(2023){\c{C}}akıro{\u{g}}lu, Island, Xie, Frisenda, and Castellanos-Gomez]{Cakroglu2023AnMaterials}
{\c{C}}akıro{\u{g}}lu,~O.; Island,~J.~O.; Xie,~Y.; Frisenda,~R.; Castellanos-Gomez,~A. {An Automated System for Strain Engineering and Straintronics of 2D Materials}. \emph{Advanced Materials Technologies} \textbf{2023}, \emph{8}, 2201091\relax
\mciteBstWouldAddEndPuncttrue
\mciteSetBstMidEndSepPunct{\mcitedefaultmidpunct}
{\mcitedefaultendpunct}{\mcitedefaultseppunct}\relax
\EndOfBibitem
\bibitem[Harats \latin{et~al.}(2014)Harats, Livneh, Zaiats, Yochelis, Paltiel, Lifshitz, and Rapaport]{Harats2014}
Harats,~M.~G.; Livneh,~N.; Zaiats,~G.; Yochelis,~S.; Paltiel,~Y.; Lifshitz,~E.; Rapaport,~R. {Full spectral and angular characterization of highly directional emission from nanocrystal quantum dots positioned on circular plasmonic lenses}. \emph{Nano Letters} \textbf{2014}, \emph{14}, 5766--5771\relax
\mciteBstWouldAddEndPuncttrue
\mciteSetBstMidEndSepPunct{\mcitedefaultmidpunct}
{\mcitedefaultendpunct}{\mcitedefaultseppunct}\relax
\EndOfBibitem
\bibitem[Livneh \latin{et~al.}(2016)Livneh, Harats, Istrati, Eisenberg, and Rapaport]{Livneh2016}
Livneh,~N.; Harats,~M.~G.; Istrati,~D.; Eisenberg,~H.~S.; Rapaport,~R. {Highly Directional Room-Temperature Single Photon Device}. \emph{Nano Letters} \textbf{2016}, \emph{16}, 2527--2532\relax
\mciteBstWouldAddEndPuncttrue
\mciteSetBstMidEndSepPunct{\mcitedefaultmidpunct}
{\mcitedefaultendpunct}{\mcitedefaultseppunct}\relax
\EndOfBibitem
\bibitem[Livneh \latin{et~al.}(2015)Livneh, Harats, Yochelis, Paltiel, and Rapaport]{Livneh2015EfficientNanoantenna}
Livneh,~N.; Harats,~M.~G.; Yochelis,~S.; Paltiel,~Y.; Rapaport,~R. {Efficient Collection of Light from Colloidal Quantum Dots with a Hybrid Metal–Dielectric Nanoantenna}. \emph{ACS Photonics} \textbf{2015}, \emph{2}, 1669--1674\relax
\mciteBstWouldAddEndPuncttrue
\mciteSetBstMidEndSepPunct{\mcitedefaultmidpunct}
{\mcitedefaultendpunct}{\mcitedefaultseppunct}\relax
\EndOfBibitem
\bibitem[Kovalchuk \latin{et~al.}(2020)Kovalchuk, Harats, L{\'{o}}pez-Pol{\'{i}}n, Kirchhof, H{\"{o}}flich, and Bolotin]{Kovalchuk2020NeutralWS2}
Kovalchuk,~S.; Harats,~M.~G.; L{\'{o}}pez-Pol{\'{i}}n,~G.; Kirchhof,~J.~N.; H{\"{o}}flich,~K.; Bolotin,~K.~I. {Neutral and charged excitons interplay in non-uniformly strain-engineered WS2}. \emph{2D Materials} \textbf{2020}, \emph{7}, 035024\relax
\mciteBstWouldAddEndPuncttrue
\mciteSetBstMidEndSepPunct{\mcitedefaultmidpunct}
{\mcitedefaultendpunct}{\mcitedefaultseppunct}\relax
\EndOfBibitem
\bibitem[Schuller \latin{et~al.}(2013)Schuller, Karaveli, Schiros, He, Yang, Kymissis, Shan, and Zia]{Schuller2013OrientationNanomaterials}
Schuller,~J.~A.; Karaveli,~S.; Schiros,~T.; He,~K.; Yang,~S.; Kymissis,~I.; Shan,~J.; Zia,~R. {Orientation of luminescent excitons in layered nanomaterials}. \emph{Nature Nanotechnology 2013 8:4} \textbf{2013}, \emph{8}, 271--276\relax
\mciteBstWouldAddEndPuncttrue
\mciteSetBstMidEndSepPunct{\mcitedefaultmidpunct}
{\mcitedefaultendpunct}{\mcitedefaultseppunct}\relax
\EndOfBibitem
\bibitem[Brotons-Gisbert \latin{et~al.}(2019)Brotons-Gisbert, Proux, Picard, Andres-Penares, Branny, Molina-S{\'{a}}nchez, S{\'{a}}nchez-Royo, and Gerardot]{Brotons-Gisbert2019Out-of-planeSelenide}
Brotons-Gisbert,~M.; Proux,~R.; Picard,~R.; Andres-Penares,~D.; Branny,~A.; Molina-S{\'{a}}nchez,~A.; S{\'{a}}nchez-Royo,~J.~F.; Gerardot,~B.~D. {Out-of-plane orientation of luminescent excitons in two-dimensional indium selenide}. \emph{Nature Communications 10:1} \textbf{2019}, \emph{10}, 1--10\relax
\mciteBstWouldAddEndPuncttrue
\mciteSetBstMidEndSepPunct{\mcitedefaultmidpunct}
{\mcitedefaultendpunct}{\mcitedefaultseppunct}\relax
\EndOfBibitem
\bibitem[Novotny and Hecht(2012)Novotny, and Hecht]{Novotny2012PrinciplesNano-Optics}
Novotny,~L.; Hecht,~B. {Principles of Nano-Optics}. \emph{Principles of Nano-Optics} \textbf{2012}, \emph{9781107005464}, 1--564\relax
\mciteBstWouldAddEndPuncttrue
\mciteSetBstMidEndSepPunct{\mcitedefaultmidpunct}
{\mcitedefaultendpunct}{\mcitedefaultseppunct}\relax
\EndOfBibitem
\bibitem[Lukosz(1979)]{Lukosz1979LightOrientation}
Lukosz,~W. {Light emission by magnetic and electric dipoles close to a plane dielectric interface. III. Radiation patterns of dipoles with arbitrary orientation}. \emph{JOSA, Vol. 69, Issue 11, pp. 1495-1503} \textbf{1979}, \emph{69}, 1495--1503\relax
\mciteBstWouldAddEndPuncttrue
\mciteSetBstMidEndSepPunct{\mcitedefaultmidpunct}
{\mcitedefaultendpunct}{\mcitedefaultseppunct}\relax
\EndOfBibitem
\bibitem[Oskooi \latin{et~al.}(2010)Oskooi, Roundy, Ibanescu, Bermel, Joannopoulos, and Johnson]{Oskooi2010Meep:Method}
Oskooi,~A.~F.; Roundy,~D.; Ibanescu,~M.; Bermel,~P.; Joannopoulos,~J.~D.; Johnson,~S.~G. {Meep: A flexible free-software package for electromagnetic simulations by the FDTD method}. \emph{Computer Physics Communications} \textbf{2010}, \emph{181}, 687--702\relax
\mciteBstWouldAddEndPuncttrue
\mciteSetBstMidEndSepPunct{\mcitedefaultmidpunct}
{\mcitedefaultendpunct}{\mcitedefaultseppunct}\relax
\EndOfBibitem
\bibitem[Hentschel and Schomerus(2002)Hentschel, and Schomerus]{Hentschel2002FresnelMicroresonators}
Hentschel,~M.; Schomerus,~H. {Fresnel laws at curved dielectric interfaces of microresonators}. \emph{Physical Review E} \textbf{2002}, \emph{65}, 045603\relax
\mciteBstWouldAddEndPuncttrue
\mciteSetBstMidEndSepPunct{\mcitedefaultmidpunct}
{\mcitedefaultendpunct}{\mcitedefaultseppunct}\relax
\EndOfBibitem
\bibitem[Luhn and Hentschel(2019)Luhn, and Hentschel]{Luhn2019AnalyticalInterfaces}
Luhn,~S.; Hentschel,~M. {Analytical Fresnel laws for curved dielectric interfaces}. \emph{Journal of Optics} \textbf{2019}, \emph{22}, 015605\relax
\mciteBstWouldAddEndPuncttrue
\mciteSetBstMidEndSepPunct{\mcitedefaultmidpunct}
{\mcitedefaultendpunct}{\mcitedefaultseppunct}\relax
\EndOfBibitem
\end{mcitethebibliography}
\end{document}